\def\3he{$^3$He}
\def\4he{$^4$He}
\def\7li{$^7$Li}

\def\ga{\mathrel{\mathpalette\fun >}}
\def\fun#1#2{\lower3.6pt\vbox{\baselineskip0pt\lineskip.9pt
  \ialign{$\mathsurround=0pt#1\hfil##\hfil$\crcr#2\crcr\sim\crcr}}}
\def\beq#1\eeq{\begin{equation}#1\end{equation}}
\def\ie{{\it i.e.},}
\def\eg{{\it e.g.},}
\def\Yp{Y$_{\rm P}$}
\def\hii{H\thinspace{$\scriptstyle{\rm II}$}}
\def\hi{H\thinspace{$\scriptstyle{\rm I}$}}
\newcommand{\Deln}{\ensuremath{\Delta N_\nu}}
\newcommand{\nnu}{\ensuremath{N_\nu}}
\newcommand{\epm}{\ensuremath{e^{\pm}\;}}
\def\etal{{\it et al}}

\documentclass{PoS}
\usepackage{graphicx}

\title{Primordial Nucleosynthesis: The Predicted 
and Observed Abundances and Their Consequences}

\ShortTitle{Primordial Nucleosynthesis}

\author{\speaker{Gary Steigman}\thanks{A footnote may follow.}\\
        Departments of Physics and Astronomy, Center for Cosmology 
        and Astro-Particle Physics\\
        The Ohio State University, Columbus, OH, USA\\
        E-mail: \email{steigman@mps.ohio-state.edu}}

\abstract{For a brief time in its early evolution the Universe 
was a cosmic nuclear reactor.  The expansion and cooling 
of the Universe limited this epoch to the first few minutes,  
allowing time for the synthesis in astrophysically interesting 
abundances of only the lightest nuclides (D, \3he, \4he, \7li).  
For big bang nucleosynthesis (BBN) in the standard models of 
cosmology and particle physics (SBBN), the SBBN-predicted 
abundances depend on only one adjustable parameter, the 
baryon density parameter (the ratio by number of baryons 
(nucleons) to photons). The predicted and observed abundances 
of the relic light elements are reviewed, testing the internal 
consistency of primordial nucleosynthesis.  The consistency 
of BBN is also explored by comparing the values of the 
cosmological parameters inferred from primordial nucleosynthesis 
for the standard model and for models with non-standard early 
Universe expansion rates with those determined from studies 
of the cosmic background radiation, which provides a snapshot 
of the Universe some 400 thousand years after BBN ended.}

\FullConference{11th Symposium on Nuclei in the Cosmos, NIC XI\\
		July 19-23, 2010\\
		Heidelberg, Germany}

\begin{document}

\section{Introduction}

During its early evolution the Universe passed through a 
very brief epoch during which its high temperature and 
particle density allowed for nuclear reactions among 
nucleons building the lightest nuclides (D, \3he, \4he, 
\7li) in astrophysically interesting abundances.  As a 
result the comparison between the predicted and observed 
primordial abundances of the light nuclides provides a 
unique window on the early evolution of the Universe 
and a key probe of the standard models of cosmology 
and particle physics.  Much later in the evolution of the 
Universe, after electrons combine with protons and alphas 
at ``recombination'', the cosmic background photons are 
free to propagate, forming the cosmic microwave background 
(CMB) radiation, the black body spectrum of radiation observed 
today at $T_{0} = 2.725$~K.  Primordial, or big bang 
nucleosynthesis (BBN) and the CMB probe particle physics 
and cosmology at two very widely separated epochs in the 
evolution of the Universe: BBN when the Universe was only 
$\sim 20$ minutes old and the CMB some 400 thousand 
years later.  Comparisons of the predictions and observations 
of BBN and the CMB provide tests of the standard models of 
cosmology and particle physics and offer constraints on new 
physics.  For a recent review of these tests and constraints 
and for further references, see \cite{steigman07}.

The standard model of cosmology uses General Relativity 
to predict the evolution of an expanding Universe filled with
radiation, including three flavors of light neutrinos, and matter 
(baryons along with non-baryonic dark matter).  Neither 
non-baryonic dark matter, dark energy (or a cosmological 
constant), or spatial curvature are relevant for our discussion 
here since they play no role in the physics and evolution of 
the early Universe.  BBN in the standard model (SBBN) 
predicts the primordial abundances of the four light nuclides 
as a function of only one adjustable parameter, the ``baryon 
abundance parameter'', the ratio by number of baryons 
(nucleons) to CMB photons, $\eta_{\rm B} \equiv n_{\rm 
B}/n_{\gamma} \equiv 10^{10}\eta_{10}$.  $\eta_{\rm B}$ 
provides a measure of the baryon asymmetry of the 
Universe whose value is a goal of but, at present, not a 
prediction of models of particle physics.  A large class of 
models of cosmology and/or particle physics which go 
beyond the standard models predict a non-standard 
expansion rate during the early, radiation-dominated (RD) 
Universe.  The Hubble parameter, $H$, measures the 
universal expansion rate; for the early RD evolution the 
age of the Universe, $t$, and $H$ are related by $Ht = 1/2$.  
A non-standard expansion rate may be parameterized by 
the ``expansion rate parameter'', $S \equiv H'/H$, where 
$S = 1$ for the standard model.  Historically, a non-standard 
particle content of the early Universe has been considered 
as the source of $S \neq 1$ \cite{ssg}.  In this case, $S$ may 
be related to the ``equivalent number of additional neutrinos'', 
the ``extra'' energy density normalized to the energy density 
contributed by one flavor of standard model neutrinos: 
$\Delta$N$_{\nu} \equiv (\rho' - \rho)/\rho_{\nu}$ (N$_{\nu} 
\equiv 3 + \Delta$N$_{\nu}$).  $S$ and \Deln~provide 
equivalent parameterizations of non-standard expansion rates.
Prior to \epm~annihilation, $S^{2} = 1 + 7\Delta$N$_{\nu}/43 
= 1 + 0.163\Delta$N$_{\nu}$, while post-\epm~annihilation, 
$S^{2} = 1 + 0.134\Delta$N$_{\nu}$ (see, \eg~\cite{vimal}).  
However, keep in mind that \Deln~$\neq 0$ need not be 
the result of extra neutrino flavors but could be due to other 
extensions of the standard models.  In particular, \Deln~need 
not be an integer and need not be positive.  For example, if 
the gravitational constant were different in the early Universe 
(at BBN and/or at recombination) from its value today, $G_{\rm 
BBN}/G_{0} = 1 + 0.163\Delta$N$_{\nu}$ and $G_{\rm REC}/
G_{0} = 1 + 0.134\Delta$N$_{\nu}$.  While SBBN only depends 
on $\eta_{\rm B}$, BBN depends on $S$ (or  \Deln) as well. 

BBN at $\sim 3$~minutes and the CMB (recombination) some 
$\sim 400$~thousand years later provide complementary probes 
of the physics and early evolution of the Universe.  This review 
of BBN addresses the following questions.  Do the BBN-predicted 
abundances agree with the observationally-inferred primordial 
abundances (\ie~is BBN internally consistent)?  Do the values 
of $\eta_{\rm B}$ inferred from BBN and the CMB agree?  Do 
the values of \Deln~derived from BBN and the CMB agree 
and, are \Deln(BBN) = \Deln(CMB) = 0?

\section{SBBN-Predicted And Observationally-Inferred 
Primordial Abundances}
\label{sbbn}

The observations of value in inferring the primordial abundances 
are diverse, from the oldest, most metal-poor stars in the Galaxy 
(\7li), to \hii~regions in the Galaxy (\3he) and extragalactic \hii~regions 
(\4he), to cool, neutral (\hi) gas in the Lyman-alpha forest (D).  To a 
greater or lesser extent, the material in these regions have experienced 
some post-BBN nuclear processing, potentially modifying the original 
relic abundances.  The post-BBN evolution of deuterium is simple 
and monotonic: as gas is cycled through stars D is burned to \3he 
and beyond \cite{els}.   As a result, the deuterium abundance 
observed anywhere, at any time in the evolution of the Universe 
provides a {\it lower} bound to the primordial value and, the 
observed deuterium abundance should reveal a plateau at the 
primordial value in regions at high redshift and/or low metallicity 
where minimal stellar processing has occurred.  In contrast, the 
post-BBN evolution of \3he is considerably more complicated.  
While {\it some} of the \3he incorporated into stars (along with 
any D which is burned to \3he) survives in the cooler, outer 
regions, most is burned away in the hotter interiors and, newly 
synthesized \3he is produced by incomplete hydrogen burning 
in some stars.  The net effect of these competitive processes 
depends on detailed models of stellar structure and evolution, 
along with models of galactic chemical evolution.  If there is 
net production of \3he in the course of chemical evolution, the 
\3he abundance should correlate with the ``metal'' (\ie~CNO...) 
abundance which, in turn, is correlated with location in the Galaxy; 
for net destruction an anticorrelation would be expected.  Since 
stars burn hydrogen to helium in the course of their evolution, 
the post-BBN evolution of \4he is simple and monotonic (similar 
to that of deuterium).  The observationally-inferred abundance of 
\4he should correlate with metallicity (\ie~CNO...), extrapolating 
to the primordial value at zero metallicity.  Finally, like deuterium, 
most lithium is destroyed when gas is cycled through stars.  
However, some stars appear to be net producers of \7li and 
collisions between cosmic ray and interstellar nuclei also 
produce post-BBN lithium.  All these effects must be kept 
in mind when using the observations to infer the primordial 
abundances.

\subsection{Deuterium}

Given its simple post-BBN evolution and the sensitivity of the 
BBN-predicted abundance to the baryon abundance parameter 
((D/H)$_{\rm DP} \propto \eta_{\rm B}^{-1.6}$, so that a $\sim 
10\%$ determination of (D/H)$_{\rm P}$ yields a $\sim 6\%$ 
measurement of $\eta_{\rm B}$), deuterium is the baryometer 
of choice.  There is, unfortunately, a very small set of D abundances 
inferred from the spectra of only seven, high-redshift, low-metallicity, 
QSO Absorption Line Systems \cite{pettini} (the observations require 
high resolution spectra on the world's largest telescopes).  Even 
more problematic, the seven data points exhibit an unexpectedly 
large dispersion (given the quoted errors).  These variations in D 
abundances show no obvious correlation with redshift or metallicity.  
From their data Pettini {\it et al.} \cite{pettini} derive log $y_{\rm 
DP} \equiv 5 + $log(D/H)$_{\rm P} = 0.45 \pm 0.03$, where the 
error has been inflated from the error in the mean to account for 
the large dispersion.  In the context of SBBN this estimate of the 
primordial D abundance corresponds to  a baryon abundance of 
$\eta_{10}(\rm SBBN) = 5.80 \pm 0.27$.

\subsection{Helium-3}

For the baryon abundance parameter inferred from SBBN using the 
Pettini {\it et al.} deuterium abundance, the SBBN-predicted \3he 
abundance is $y_{3\rm P}$(SBBN)~$ \equiv 10^{5}$(\3he/H)$_{\rm P} 
= 1.08 \pm 0.04$.  Aside from the solar system \cite{gg}, \3he is 
observed in Galactic \hii~regions via the spin-flip transition (the 
analog of the 21 cm line in neutral hydrogen) in singly-ionized \3he.  
The most complete data set is that of Bania, Rood \& Balser \cite{brb}.  
The gas in these Galactic \hii~regions has been processed through 
several generations of stars so, not surprisingly, the \3he abundances 
reveal a large variation, likely due to post-BBN production/destruction.  
However, these \3he~abundances show no clear correlation with either 
metallicity (\eg~O/H) or location in the Galaxy (see, \eg~Figure 9 and 
the discussion in Steigman 2007 \cite{steigman07}).  The lowest \3he 
abundances, $y_{3} \geq 1.1 \pm 0.2$, which are adopted by Bania, 
Rood \& Balser as an estimate of (or, a lower bound to) the primordial 
abundance, are in excellent agreement with the SBBN-predicted 
abundance, providing support for the internal consistency of SBBN 
(two predicted abundances, one free parameter).  However, the 
lack of a correlation between the \3he abundances and metallicity 
or location in the Galaxy is puzzling.

\subsection{Helium-4}

\begin{figure} 
\begin{center}
\includegraphics[width=.6\textwidth]{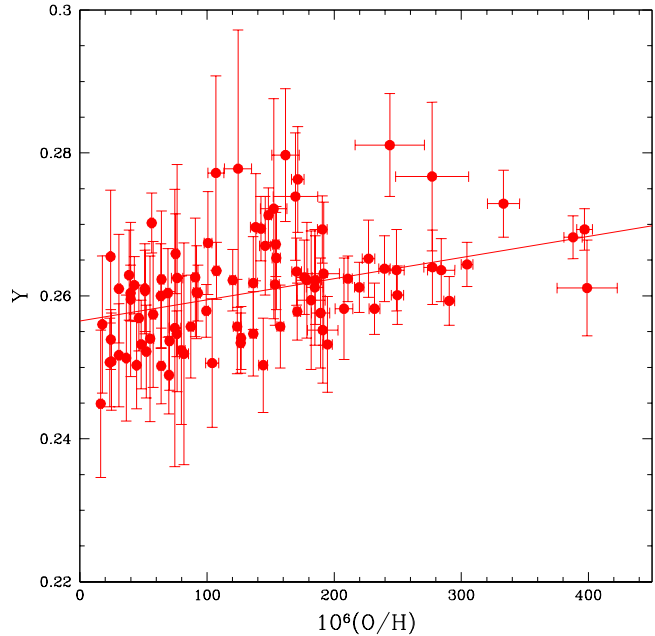} 
\end{center}
\caption{The Izotov \& Thuan 2010 (IT10) \cite{it10} 
helium and oxygen abundances.  The solid line is the 
IT10 best fit for a {\it linear} Y versus O/H relation; 
see the text.} 
\label{hevsoit10} 
\end{figure} 

As the second most abundant nuclide in the Universe \4he is 
observed in many astrophysical environments.  The most useful 
observations for BBN are those of the hydrogen and helium 
recombination lines from low-metallicity, extragalactic \hii~regions.  
This is a field with a long history; see \cite{steigman07} for 
references.  With the presently available large data set it 
has become increasingly clear that systematic uncertainties 
(which are always difficult to quantify) dominate the statistical 
errors.  Recently, Izotov \& Thuan \cite{it10} used their data 
of 93 spectra for 86 low-metallicity extragalactic \hii~regions
to derive the helium and oxygen abundances shown in Figure
\ref{hevsoit10}.  A linear fit to the Y -- O/H data extrapolated 
to zero metallicity yields Y$_{\rm P} = 0.2565 \pm 0.0010 
(stat) \pm 0.0050 (syst)$ \cite{it10}.  For the comparison here 
with BBN I combine the errors {\it linearly}, adopting Y$_{\rm P} 
= 0.2565 \pm 0.0060$.  These results are consistent with the 
recent determinations of \Yp~from a much more limited data 
set ($\leq 9$  \hii~regions) by Aver, Olive \& Skillman \cite{aos}.  
For SBBN with the D-determined baryon density parameter, 
Y$_{\rm P}$(SBBN)$~= 0.2482 \pm 0.0007$.  While the 
central value of the SBBN-predicted abundance appears 
to be in conflict with the observationally-inferred value, 
given the relatively large uncertainty in the latter estimate, 
the difference is only at the $\sim 1.4 \sigma$ level.   SBBN 
is consistent with the observationally-inferred primordial 
abundances of D, \3he, and \4he (within the errors).

\subsection{Lithium-7}

\begin{figure} 
\begin{center}
\includegraphics[width=.6\textwidth]{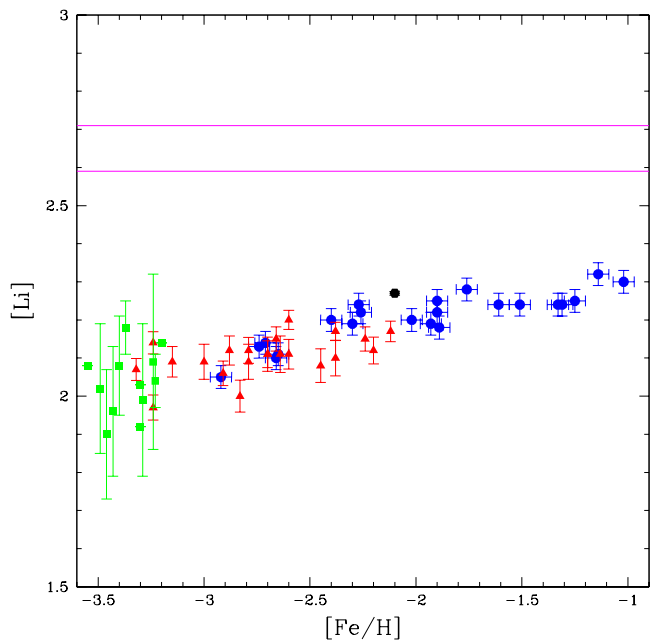} 
\end{center}
\caption{The lithium abundances ([Li]~$\equiv 12 + $log(Li/H))
as a function of metallicity ([Fe/H]) as derived from observations 
of very metal-poor stars in the Galaxy.  See the text for references.  
The horizontal band shows the $\pm 1\sigma$ SBBN prediction.} 
\label{livsfe} 
\end{figure} 

For the D-determined baryon density parameter the SBBN-predicted
lithium abundance is [Li]$_{\rm SBBN} \equiv 12 + $log(Li/H)$_{\rm 
SBBN} = 2.65 \pm 0.06$.  In Figure \ref{livsfe} are shown the lithium
abundances derived from observations of the oldest, most metal-poor
stars in the Galaxy \cite{boesgaard05,asplund06,aoki09,lind09}.  
These stars should provide a sample of the ``lithium plateau'', the 
nearly primordial abundance in systems which have experienced 
very little stellar processing.  It is difficult to identify a lithium plateau 
from the data shown in Figure \ref{livsfe} and it is clear that {\bf none} 
of the observationally-inferrred lithium abundances is even close to 
the SBBN-predicted value.  The difference is $\sim 0.5 - 0.6$~dex 
or, a factor of 3 -- 4.  Lithium is a problem.  However, since the 
target stars have had  $\ga 10$~Gyr to modify their surface 
abundances by mixing with the lithium-depleted interior material, 
it is unclear if the problem is one for cosmology or particle 
physics or, for stellar astrophysics.

\subsection{Summary for SBBN}

The observationally-inferred primordial abundances of the light nuclides
are subject to difficult to quantify systematic uncertainties along with 
the usual statistical errors.  The fact that D, \3he, \4he, and \7li are 
observed by different astronomical techniques in different targets ensures 
that their derived abundances are not affected by the same systematics.
Within the errors, the SBBN-predicted and observationally-inferred 
primordial abundances of D, \3he, and \4he are in good agreement,
providing support for the standard models of particle physics and
cosmology (\eg~\Deln~= 0).  However, lithium is a problem.  Since
non-standard BBN offers a second free parameter, $S$ or \Deln, it is 
of interest to see how the BBN abundances constrain the combination 
of $\eta_{\rm B}$ and \Deln~and, if this additional freedom can help to 
ameliorate the lithium problem. 

\section{Non-Standard BBN}

\begin{figure} 
\begin{center}
\includegraphics[width=.6\textwidth]{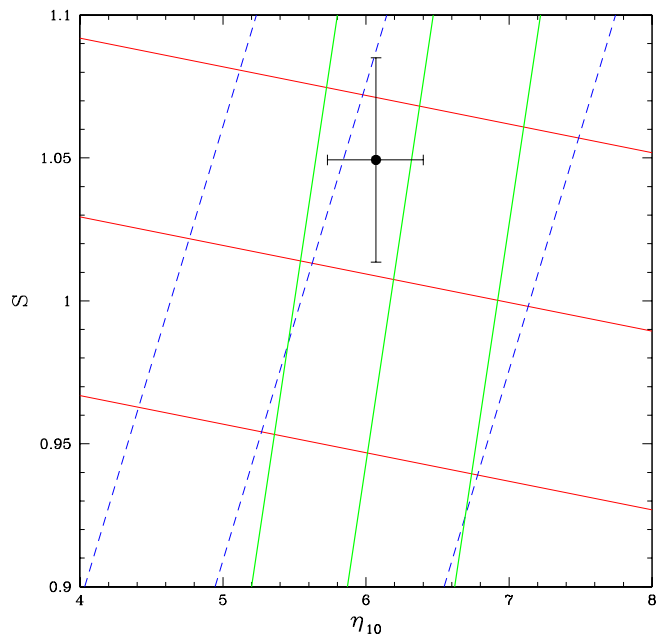} 
\end{center}
\caption{The BBN isoabundance contours in the $S$ 
versus $\eta_{10}$ plane \cite{ks,steigman07} for 
$y_{\rm DP} \equiv 10^{5}$(D/H) (dashed, blue) and 
Y (solid, red).  $y_{\rm DP}$ decreases from left to 
right from 4.0 to 3.0 to 2.0 respectively.  Y increases 
from bottom to top, from 0.24 to 0.25 to 0.26 respectively.  
Also shown are the isoabundance contours (solid, green) 
for [Li] $\equiv 12 + $log(Li/H).  [Li] increases from left 
to right from 2.6 to 2.7 to 2.8 respectively.  The data 
point with error bars corresponds to the adopted D 
and $^{4}$He abundances; see the text.
} 
\label{svsetalia} 
\end{figure} 

For the more general BBN case the relic abundances 
depend on two adjustable parameters, the baryon
density parameter $\eta_{\rm B}$ and the expansion
rate parameter $S$ (or, equivalently, \Deln).  In Figure 
\ref{svsetalia} are shown the D and \4he isoabundance 
contours, $y_{\rm DP}$ (dotted, blue) and \Yp~(solid, 
red) in the $S - \eta_{10}$ plane \cite{ks,steigman07}.  
The point with the error bars represents the 
observationally-inferred D and \4he abundances from 
\S\ref{sbbn}.  Since the $y_{\rm DP}$ and \Yp~contours 
form a grid in the $S - \eta_{10}$ plane, the D and \4he 
abundances constrain $S$ (or, \Deln) and $\eta_{10}$, 
as may be seen from Figure \ref{svsetalia}.  The adopted 
D and \4he abundances predict $\eta_{10}($BBN$) = 
6.07 \pm 0.33$ and \Deln~$= 0.62 \pm 0.46~(= 0$ at 
$\sim 1.3\sigma)$.  

Also shown in Figure \ref{svsetalia} (solid, green) 
are the \7li isoabundance contours.  The D and \4he 
constrained, BBN-predicted lithium abundance with 
both $\eta_{\rm B}$ and $S$ (\Deln) as free parameters 
yields [Li]$_{\rm BBN} = 2.66 \pm 0.06$.  Lithium is 
still a problem.

\begin{figure} 
\begin{center}
\includegraphics[width=.6\textwidth]{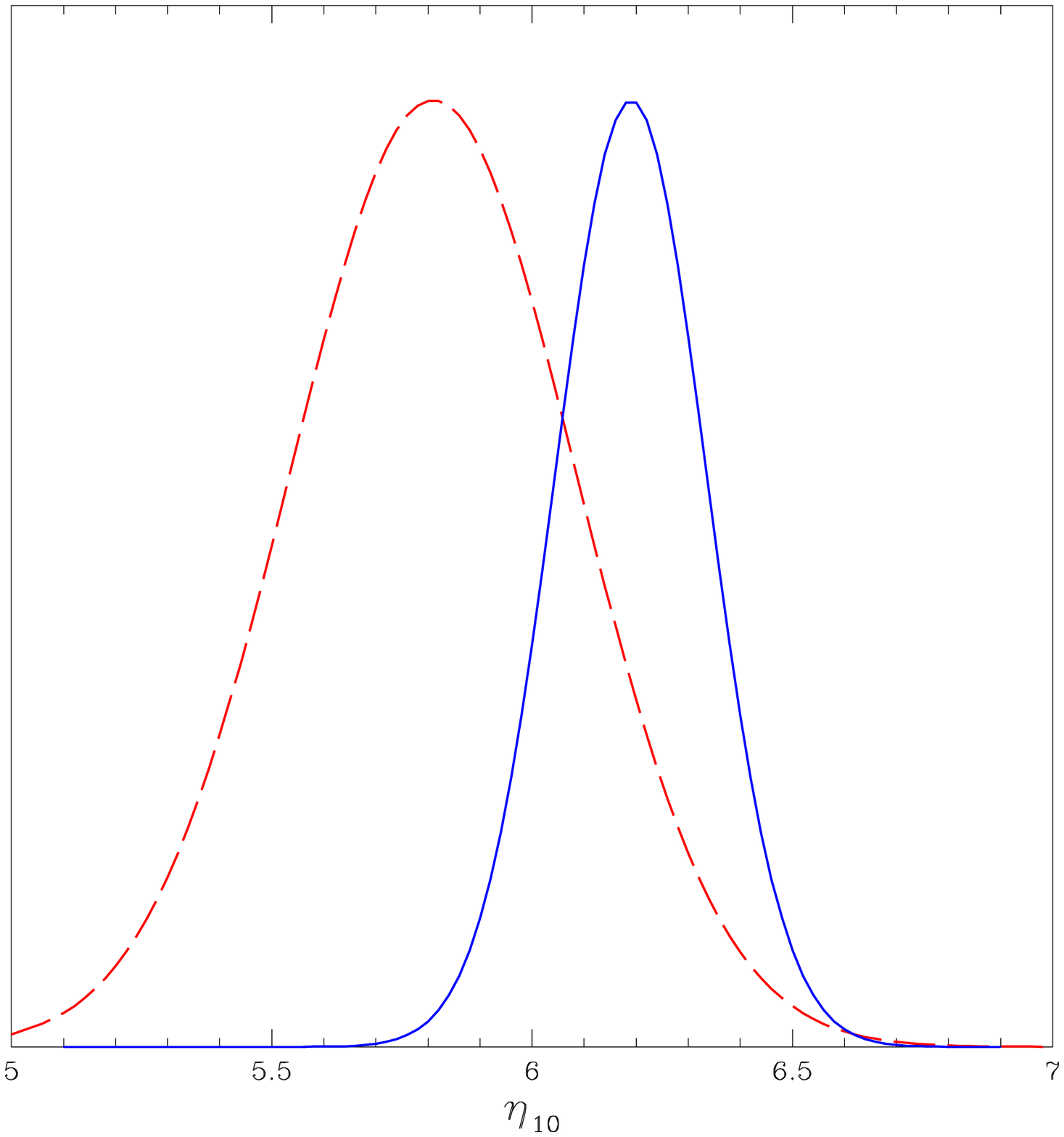} 
\end{center}
\caption{The likelihood distributions for $\eta_{10}$ 
inferred from SBBN (dashed, red) and the CMB 
(solid, blue).} 
\label{eta10} 
\end{figure} 

\begin{figure} 
\begin{center}
\includegraphics[width=.6\textwidth]{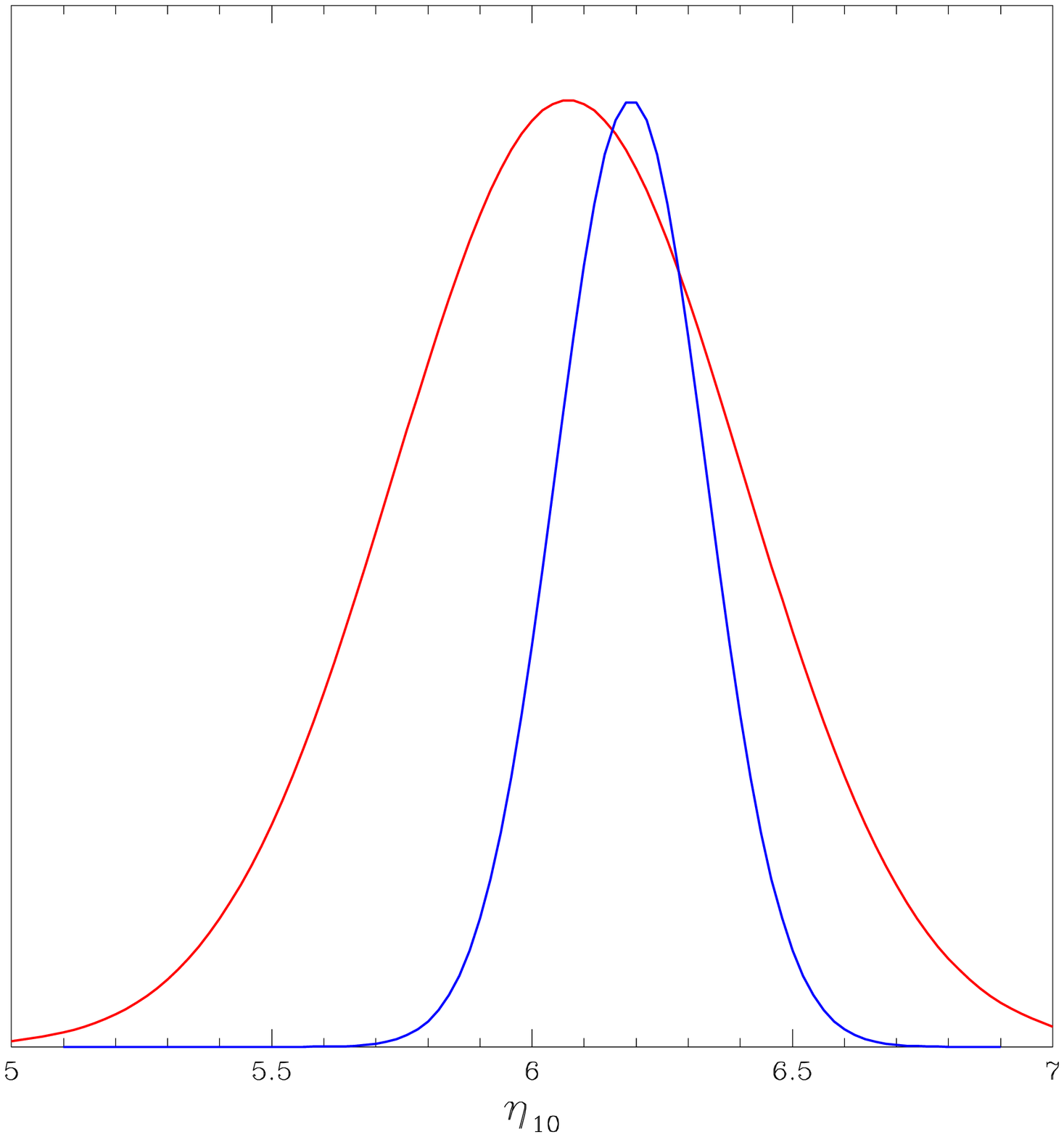} 
\end{center}
\caption{The likelihood distributions for $\eta_{10}$ 
inferred from BBN (red) and the CMB (blue).} 
\label{eta10a} 
\end{figure} 

\section{Comparison Between BBN And The CMB}

The CMB offers a probe of the later, but still early,
evolution of the Universe which complements that
from BBN.  In particular, the CMB temperature
anisotropy spectrum is sensitive to the values at
recombination of both the baryon abundance
parameter and the expansion rate parameter
(or, \Deln).

From the 7-year WMAP data Komatsu \etal~\cite
{komatsu} find $\eta_{10} = 6.190 \pm 0.145$.  If 
it is assumed that \nnu~= 3 (SBBN), for which 
$\eta_{10}({\rm SBBN}) = 5.80 \pm 0.27$, then SBBN 
and the CMB agree to within $\sim 1.3\sigma$ on 
the value of the baryon density parameter at a few 
minutes and some 400 thousand years later.  The
comparison of the SBBN and CMB inferred values of 
$\eta_{10}$ are shown in the likelihood distributions
of Figure \ref{eta10}.  However, Komatsu \etal~find 
some evidence (at the $\sim 1.5\sigma$ level) in 
support of \Deln~$\neq 0$.  For BBN (\Deln~$\neq
0$), $\eta_{10}({\rm BBN}) = 6.07 \pm 0.33$, which
is in even better agreement with CMB result for 
$\eta_{10}$, as may be seen in Figure \ref{eta10a}.

\begin{figure} 
\begin{center}
\includegraphics[width=.6\textwidth]{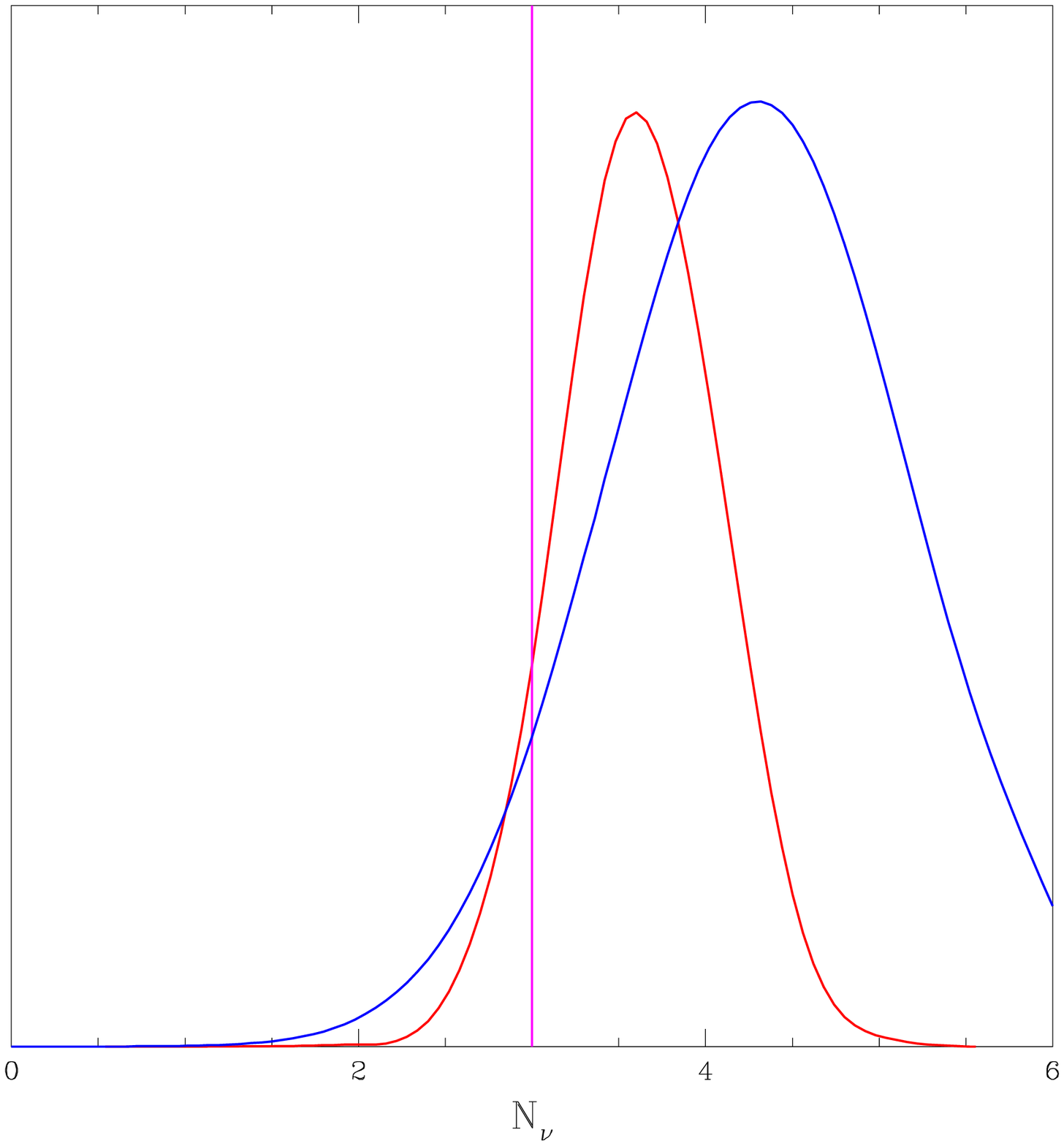} 
\end{center}
\caption{The likelihood distributions for N$_{\nu} 
\equiv 3 + \Delta$N$_{\nu}$ inferred from BBN 
(red) and the CMB (blue).  The vertical line 
(magenta) is for the standard model at 
N$_{\nu} = 3$.} 
\label{nnu10a} 
\end{figure}

For BBN and the observationally-inferred D and \4he 
relic abundances, \Deln~$ = 0.62 \pm 0.46$, which is 
consistent with \Deln~= 0 at $\sim 1.3\sigma$.  From 
the imprint on the CMB at recombination Komatsu 
\etal~\cite{komatsu} find \Deln~$ = 1.30 \pm 0.87$ 
(see also \cite{steigman10}), which differs from \Deln~= 
0 by $\sim 1.5\sigma$.  The very good agreement 
between BBN and the CMB, \Deln(CMB) -- \Deln(BBN) 
= $0.68 \pm 0.98$, as well as their overlap with \Deln~= 
0, is illustrated by the likelihood distributions in Figure 
\ref{nnu10a}.

\section{Discussion}

The predictions of SBBN (\Deln~= 0) are consistent, 
within the errors, with the observationally-inferred relic 
abundances of D, \3he, and \4he, as well as with the 
value of the baryon density parameter derived from 
the CMB.  But, lithium is a problem since its SBBN
predicted abundance differs from that derived from
observations of very metal-poor halo and globular
cluster stars by a factor of three or more.  The problem
could be in the stars (\eg~depletion or dilution of 
surface lithium during the lifetime of the oldest stars 
in the Galaxy) or in the cosmology (\eg~late-time or 
renewed BBN initiated by the decay of long-lived 
massive particles).

For non-standard BBN with \Deln~$\neq 0$ comparison 
between the predicted and observed abundances
of D and \4he constrains the allowed values of the
baryon density parameter ($\eta_{\rm B}$) and the 
expansion rate parameter ($S$) when the Universe 
was only a few minutes old.  The BBN results are 
consistent with the CMB inferred values of these 
parameters at recombination, some 400 thousand 
years later.  But, even in this case lithium remains 
a problem.  Lithium aside, the very good agreement 
between these independent probes of cosmology 
during widely separated epochs in the evolution of 
the Universe validates using the combined constraints 
from BBN and the CMB to probe non-standard 
models of cosmology and particle physics.

\subsection{Comparing The Universe At BBN And At 
Recombination} 

During most epochs in the evolution of the Universe
entropy is conserved.  The number of CMB photons in 
a comoving volume provides a measure of the entropy 
in that comoving volume.  A comparison of the number 
of photons in the comoving volume at BBN and at 
recombination provides a test of entropy conservation.  
Of course, it is necessary to define the size of the 
comoving volume, which can be accomplished by
identifying the number of baryons (nucleons) in it,
so that N$_{\gamma} = $~N$_{\rm B}/\eta_{\rm B}$.
Given the very good constraints on baryon non-conservation,
the number of baryons in a comoving volume is
(should be!) preserved from BBN to recombination,
so that a comparison of $\eta_{\rm B}$ derived 
from BBN and from the CMB constrains any 
entropy production in the intervening epochs: 
N$_{\gamma}$(REC)/N$_{\gamma}$(SBBN) =$~0.94 
\pm 0.05$ and N$_{\gamma}$(REC)/N$_{\gamma}$(BBN) 
=$~0.98 \pm 0.06$.

Similarly, the combined results of BBN and the CMB may be 
used to compare the value of the gravitational constant, $G$, 
at BBN and at recombination with each other and with the 
present value ($G_{0}$).  Since $G_{\rm BBN}/G_{0}  = 1 
+ 0.163\Delta$N$_{\nu}({\rm BBN})$ and $G_{\rm REC}/G_{0} 
= 1 + 0.134\Delta$N$_{\nu}({\rm REC})$ \cite{vimal}, the 
values of \Deln~at BBN and at recombination constrain 
the early Universe strength of gravity, leading to: $G_{\rm 
REC}/G_{\rm BBN} = 1.07 \pm 0.13$, $G_{\rm BBN}/G_{0} 
= 1.10 \pm 0.07$ and, $G_{\rm REC}/G_{0}  = 1.17 \pm 0.12$.

As a last example of the value of combining the constraints from 
BBN and the CMB consider the effect of a massive particle which 
decays after BBN but before recombination.  If the decay occurs 
too late for the relativistic decay products to be thermalized 
(producing extra CMB photons whose presence has already 
been constrained above), the energy they carry will, nevertheless, 
contribute to the total energy density when the Universe is radiation 
dominated.  Since $(\rho'_{\rm R}/\rho_{\rm R})_{\rm BBN} = 1 + 
0.163\Delta$N$_{\nu}({\rm BBN})$ and $(\rho'_{\rm R}/\rho_{\rm 
R})_{\rm REC} = 1 + 0.134\Delta$N$_{\nu}({\rm REC})$, 
\Deln~measures the presence of ``extra'' relativistic energy.  
Comparing \Deln~at BBN with its value at recombination it is 
seen that they agree, $(\rho'_{\rm R}/\rho_{\rm R})_{\rm BBN} 
= (\rho'_{\rm R}/\rho_{\rm R})_{\rm REC}$ within $0.5\sigma$.

\section{Conclusions}

For \nnu~$\approx 3$ BBN is internally consistent
(but lithium is a problem!) and is in agreement with
the CMB.  The very good agreement between BBN
and the CMB permits their use in combination to
constrain some examples of non-standard models of 
cosmology and particle physics.  While celebrating
the success of BBN it should be kept in mind that
some challenges remain.  For example, why is the 
dispersion among the observed deuterium abundances
so large?  Or, why don't the \3he abundances observed
in Galactic \hii~regions correlate with the oxygen
abundances or with location in the Galaxy?  And,
for \4he, the second most abundant element in the
Universe, how large are the systematic errors in its
observationally-inferred primordial abundance and,
are there observing strategies to reduce or eliminate 
at least some of them?  This active research area of 
importance for cosmology and particle physics would 
benefit greatly from more data.

\acknowledgments

I am pleased to acknowledge informative discussions
and correspondence with R.~Cyburt, G.~Ferland, Y.~Izotov, 
A.~Korn, and T.~Prodanovi\'{c}.  The research reported 
here is supported at The Ohio State University by a grant 
from the U. S. Department of Energy.

\end{document}